# Probability distribution in the Toda system: The singular route to a steady state

Srdjan Petrović, Nikola Starčević

Laboratory of Physics, Vinča Institute of Nuclear Sciences, P. O. Box 522, 11001 Belgrade, Serbia

Nace Stojanov

Institute of Physics, Faculty of Natural Sciences and Mathematics, Ss. Cyril and Methodius University, Arhimedova 3, 1000 Skopje, North Macedonia

Liang Huang

Lanzhou Center for Theoretical Physics, Key Laboratory of Quantum Theory and Applications of MoE, Lanzhou University, Lanzhou, Gansu 730000, China

**Abstract**

This study reports on the evolution of the probability distribution in the configuration space of the two-dimensional Toda system. The distribution is characterized by singularities, which predominantly take two forms: double-cusped triangular lines and lines parallel to the equipotential line that defines the accessible region. Over time, the number of these singular patterns increases linearly. Consequently, at very large times, the singular patterns fully occupy the accessible area, resulting in a steady state probability distribution with a pronounced singular peak at the center.

Changes in the singular patterns arise solely from the system's intrinsic dynamics rather than variations in its parameters, emphasizing the system's self-organizing nature over time. These results provide a deeper understanding of the collective motion of particles in symmetric, bounded, two-dimensional conservative systems.

**Introduction**

We study collective motion in the Toda lattice [1] applying a numerical experiment method. The famous Fermi, Pasta, and Ulam (FPU) problem [2] investigating the collective motion of the lattice, assuming a nonlinear perturbation, was one of the first examples of using a numerical experiment (in the very beginning of the era of computer development) and an inspiration for solving many different problems in nonlinear dynamics associated with the FPU system [3]. Hénon and Heiles studied the motion of stars in a galaxy with cylindrical symmetry by reducing the problem to a two-dimensional (2D) mean-filed potential [4]. Their work was pioneering in discovering chaotic motion in a conservative 2D bounded system occurring for large enough particle energies. The monography by T. P. Weissert serves as an interesting work on the history and important contribution of numerical experiments for nonlinear systems [5].

M. Toda proposed a nonlinear exponential interaction in the lattice system, anticipating that this system could, in principle, be solved analytically [1, 6]. Following a numerical experiment by J. Ford et al. [7], which indicated that the Toda lattice is "a rare jewel in physics" - a physically interesting nonlinear integrable system - M. Hénon found analytical expressions for all independent integrals of motion, proving the integrability of the Toda lattice [8].

An independent line of work [9], including the Toda continuum lattice case [6, 10], leads to the formulation of soliton physics. A comprehensive monograph on integrable continuous systems and solitons, with a chapter dedicated to the Toda lattice, was written by L. D. Faddeev and L. A. Takhtajan [11]. The significance of the Toda lattice in the field of integrable dynamical systems can be found in a recent review [12].

This study explores the collective motion and global characteristics of non-interacting particles in the 2D Toda system through the probability distribution in configuration space. The concept of the probability distribution in configuration space provides fundamental insights into the collective behavior of diverse systems, including astrophysical [13], biological [14], and neural network [15] systems, among others.

The analytical approach for studying probability distributions is based on statistical physics methods [16]. As a result, the Fokker-Planck equation can be derived to describe a probability distribution dynamic [16, 17]. When applied to a system in configuration space, this equation governs the evolution of the probability distribution of a system, incorporating both deterministic and stochastic (diffusion) terms.

For a conservative Hamiltonian system relevant to this work, the stochastic (diffusion) term vanishes, reducing the Fokker-Planck equation to the probability continuity equation, with the deterministic term depending on the gradient of the potential. In this case, analytical solutions can be obtained for only the simplest systems, while numerical solutions cannot capture all the details of the distribution due to approximations made during the equation's derivation. In contrast, we construct the probability function directly from first principles, without any approximations, using individual particle trajectories. This approach is much more accurate and will be explained in greater detail in the subsequent sections.

Our analysis of the probability distribution focuses on the singularities of the 2D mapping in configuration space induced by the Hamiltonian flow. These singularities arise when the mapping becomes non-invertible. A central problem in singularity theory is the study of structurally stable singularities - those that preserve their topological characteristics under small perturbations [18]. A significant advancement in this field was the investigation of singular properties of equilibrium sets of function families (when the function's gradient vanishes), which led to the development of catastrophe theory [19, 20]. Parallel to this, a related research area, pioneered by A. Andronov and L. Pontryagin [21], explores stability and abrupt behavioral changes in dynamical systems under small variations in system parameters. This work laid the foundation for bifurcation theory, which has found broad applications [22, 23].

It is important to emphasize that our study examines the probability distribution and their singular patterns as they evolve continuously over time. In this context, changes in these patterns emerge from the system's intrinsic dynamics rather than external parameter variations. Our aim is to compare probability distribution with its associated singular patterns and investigate the role these patterns play in the evolution of probability distribution.

This work has been inspired by the work of J. Ford et al. [7], extending their findings to the collective behavior of non-interacting particles in the 2D Toda system. Additionally, it is a continuation of the author's previous work on the scattering of protons through very long chiral carbon nanotubes, in which the rainbow route to equilibration was presented [24]. That system was based on an effectively one-dimensional potential.

**Theory**

The motion of particles with the same unit masses in the three-dimensional Toda lattice with the periodic boundary conditions can be described by the following Hamiltonian [7]:

$$H = \frac{1}{2}\sum_{i=1}^{3} P_i^2 + e^{-(Q_1-Q_3)} + e^{-(Q_2-Q_1)} + e^{-(Q_3-Q_2)} - 3, \tag{1}$$

where $Q_i$ and $P_i$, $i = 1, 2, 3$ are dimensionless conjugate variables.

After the appropriate canonic transformation, considering that there is a constant of motion associated with the uniform translation of the lattice, the corresponding 2D Hamiltonian reads [7]:

$$H = \frac{1}{2}(p_x^2 + p_y^2) + V(x, y), \tag{2}$$

where,

$$V(x, y) = \frac{1}{24}(e^{2x\sqrt{3}+2y} + e^{-2x\sqrt{3}+2y} + e^{-4y}) - \frac{1}{8}, \tag{3}$$

is the expression for the 2D Toda potential, assuming standard notation for $x$ and $y$ as coordinates in the configuration space. $p_x$ and $p_y$ are their conjugated impulse coordinates, respectively. In the following text, expression (3) will be referred to as the Toda potential.

Fig. 1(a) shows the equipotential lines for the Toda potential at the following values equal to $10^{-3}, 10^{-2}, 10^{-1}, 1, 10^1, 10^2, 10^3$. It is evident that the Toda potential has a $C_{3v}$ symmetry – the symmetry of an equilateral triangle - forming a bounded potential well with one fixed stable point at the center and exhibiting exponential growth of the potential in all directions from the center.

The probability distribution for finding particles in the configuration space can be numerically defined by the expression, $\rho_{num}(x, y, t) = \frac{\Delta N(x,y,t)}{N_0}$, where $N_0$ is the initial number of particles and $\Delta N$ is the number of particles in a bin of the corresponding histogram whose centroid is located at the point $(x, y)$, for a particular time $t$. The dimensions of the bins are uniform and are determined by dividing the investigated area. While it can be argued that the numerical probability density is, in fact, the averaged probability density, our numerical experiment strongly suggests that the key features of the probability distribution can be obtained from $\rho_{num}(x, y, t)$.

The initial conditions for an ensemble of classical non-interacting particles moving in the Toda potential were chosen as follows. The particles were initially at rest, with velocities equal to zero, and uniformly distributed in the region where the Toda potential satisfies $V(x, y) < 1$. These initial conditions were imposed in analogy with a uniform parallel beam in a scattering process [24, 25]. However, they can also represent physical cases when non-interacting particles move in a 2D mean-field potential. In such cases, the initial kinetic energies of the particles are negligible compared to their potential energies, and the initial distribution in the configuration space can be approximated as uniform. Since the system described by equation (2) is conservative, the accessible area for a particle's motion is bounded by the equipotential line equal to 1 (depicted in Fig. 1(a)).

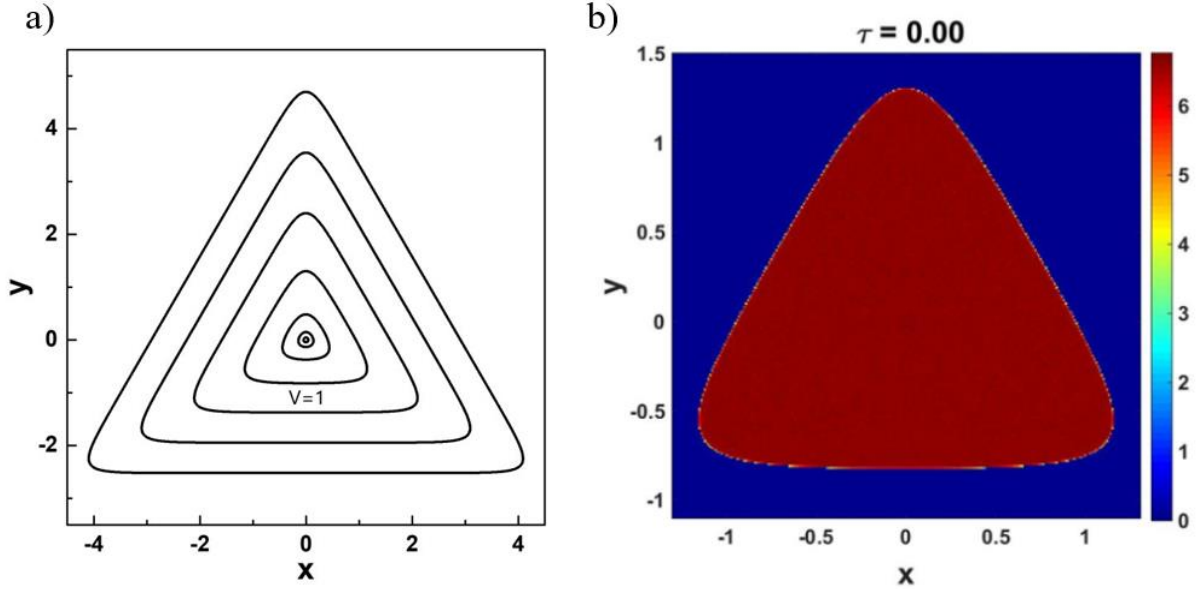

**Figure 1.** (a) Equipotential lines for the Toda potential at values equal to $10^{-3}$, $10^{-2}$, $10^{-1}$, 1, $10^1$, $10^2$, $10^3$, (b) The initial 2D distribution for $V < 1$. The jet color map in a logarithmic scale is used.

Our aim is to study the evolution of probability distribution over time, including "very large times". To define "very large times", a reference time scale must be established. By expanding the potential (3) around the center: $V(x, y) = \frac{1}{2}(x^2 + y^2) + (x^2 y - \frac{1}{3} y^3) + \cdots$, it can be seen that, near the center, the potential $V$ can be approximated by the parabolic (harmonic) polynomial, $V_0 = \frac{1}{2}(x^2 + y^2)$. Thus, time can be expressed in units of the period of harmonic motion around the center, which equals $2\pi$. This rescaled time is defined as $\tau = \frac{1}{2\pi} t$, where $t$ represents the evolution time of the system described by the Hamiltonian (2).

One can introduce the 2D Hamiltonian flow in configuration space, described by the mapping:

$$M(x_0, y_0; x, y, \tau): x_0, y_0 \to x(\tau), y(\tau), \qquad (4)$$

where $x_0, y_0$ are the initial coordinates, and $x(\tau), y(\tau)$ are the coordinates evolving from the initial coordinates at a particular time, $\tau$.

The Jacobian of the mapping (4) reads:

$$J(x_0, y_0; x, y, \tau) \equiv \begin{vmatrix} \frac{\partial x(\tau)}{\partial x_0} & \frac{\partial x(\tau)}{\partial y_0} \\ \frac{\partial y(\tau)}{\partial x_0} & \frac{\partial y(\tau)}{\partial y_0} \end{vmatrix}. \qquad (5)$$

The mapping $M$ is singular if $J = 0$. In the case of the potential $V_0$, the analytical solutions of the equations of motion can be easily obtained:

$$x(\tau) = x_0 \cos(2\pi\tau) \text{ and } y(\tau) = y_0 \cos(2\pi\tau); \frac{dx(\tau)}{d\tau}(0) = 0 \text{ and } \frac{dy(\tau)}{d\tau}(0) = 0. \tag{6}$$

It follows from (5) and (6) that $J = cos^2(2\pi\tau)$, and the mapping (4) is singular for $\tau = \frac{1}{4} + n\frac{1}{2}$, where $n = 0, 1, 2, ...$ At these particular times, the whole Hamiltonian flow converges at the center. This is a direct consequence of the unique feature of the linear harmonic oscillator, that the period of oscillations does not depend on the amplitude [26].

A classical and pioneering result by H. Whitney on the local structural stability of differentiable mappings from the plane to the plane, which is directly related to our study, states that the mapping is locally stable if and only if: (i) it is regular ($J \neq 0$), (ii) the singular pattern is of the fold type, or (iii) the singular pattern is of the cusp type [27]. A fold type singularity is topologically equivalent to a smooth line and a cusp type singularity to a semicubic parabola with a cusp at the origin [27, 20]. A mapping is globally stable if it is stable around all points. One of the important open questions in singularity theory is classification of globally stable mappings.

Therefore, the mapping $M$ (4) for the harmonic system (6) is globally stable, except at specific times: $\tau = \frac{1}{4} + n\frac{1}{2}$, where $n = 0, 1, 2, ...$. At these times, singularities correspond to isolated singular points. This result will be discussed in greater detail later in the text, when describing the singularities of the Toda system (2) for these specific times.

**Results**

The initial number of particles was 23,800,302. Their positions were generated using a Monte Carlo method with the random generator program, $ran1$ [28], while taking into account the symmetry of an equilateral triangle. This symmetry ensures that symmetric initial coordinates correspond to symmetric final coordinates. The dimension of a histogram bin was $S_{bin} = 0.01 \times 0.01$. This high particle count, and small bin size were chosen to enable detailed analysis of the probability distribution patterns.

Fig. 1(b) shows the initial distribution of particles within the configuration space presented in a jet-color map (with a blue color corresponding to minimal yield and a red color to maximal yield) on a logarithmic scale. In this work, the 2D distributions are presented in the same way. The analysis shows that the average value of the probabilities (in percentage) per bin was 0.00317% with the standard deviation equal to 0.000120%. On the other hand, $\frac{S_{bin}}{S} 100\% = 0.00317\%$, where $S$ is the area corresponding to $V(x, y) < 1$. Therefore, numerically speaking, the initial distribution was homogeneous.

The fourth-order Runge-Kutta method [28] with a fixed time step of 0.01 was used to compute the particle's trajectories. The analysis shows that by halving the time step, the Runge-Kutta method achieves a relative average accuracy on the order of $10^{-4}$, which is sufficient for the required numerical precision for the calculation of the probability distribution.

The calculated probability distributions for short times, $\tau = 2, 2.25, 3,$ and $3.25$ are shown in Figs. 2(a)-(d), respectively. For these times, one can observe two types of characteristic structures: (i) double cusped-like triangles and (ii) lines that are parallel to equipotential line $V = 1$. For $\tau = 2$ and 3, there are a total of 4 and 6 characteristic line-like structures, respectively. Therefore, the increase in the number of characteristic structures appears to be linear with time. On the other hand, defocusing of the probability distribution around the center can be observed for $\tau = 2$ and 3, while strong focusing occurs for $\tau = 2.25$ and 3.25, respectively.

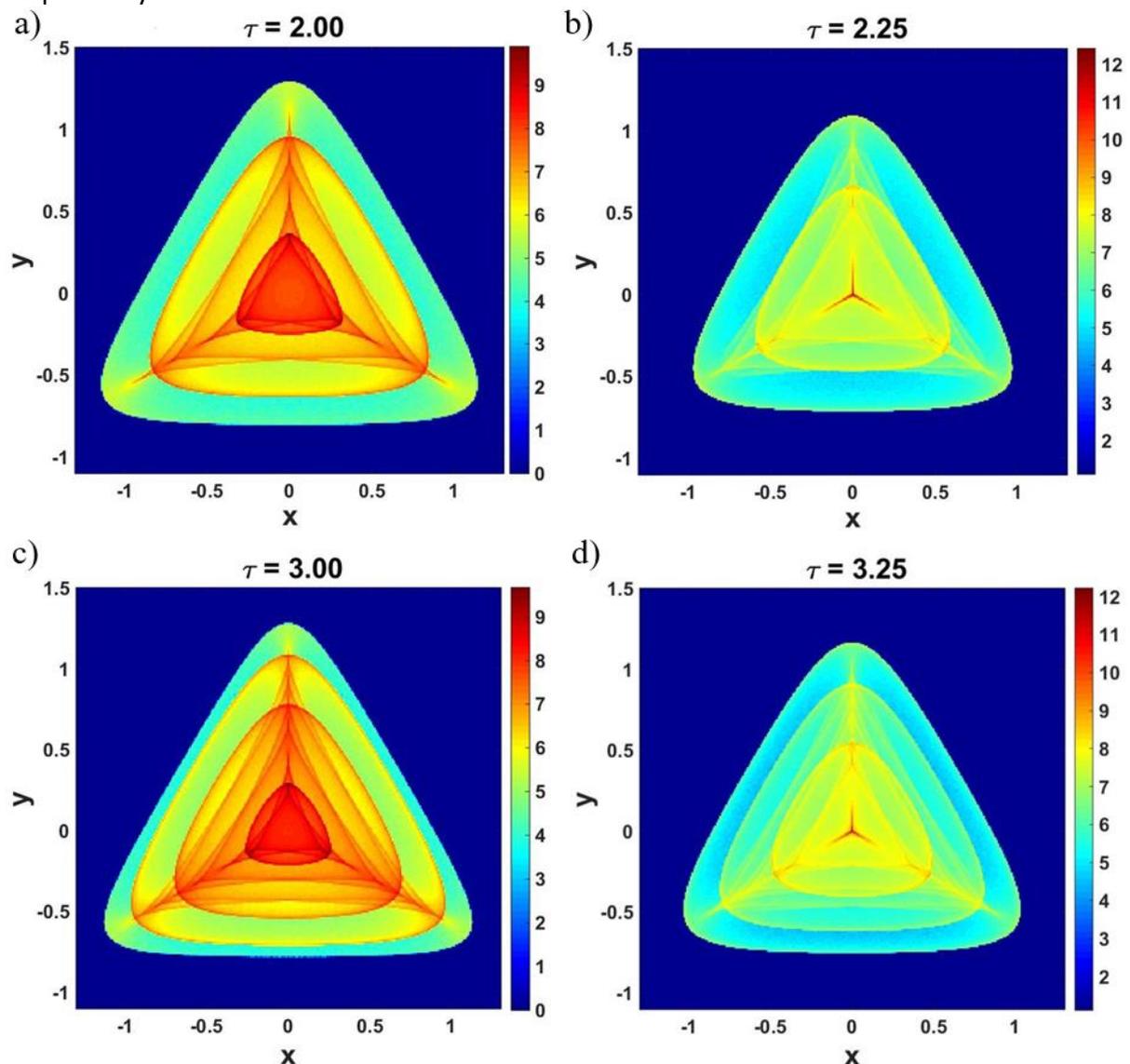

**Figure 2.** The 2D distribution at times equal to (a) $\tau = 2$, (b) $\tau = 2.25$, (c) $\tau = 3$, and (d) $\tau = 3.25$. The jet color map in a logarithmic scale is used.

Fig. 3 presents cuts through the probability distributions for $x = 0$ taken at the characteristic times corresponding to those in Fig. 2. Comparing Fig. 3 with Fig. 2 justifies the observation of the characteristic line-like structures, as one can identify peaks corresponding to the intersection of these structures with $x = 0$ axis. For $\tau = 2.25$ and 3.25, sharp, intensive peaks at the center are clearly visible. Furthermore, a trend emerges over time characterized by an increasing number of peaks, a reduction in their mutual distances, a diminishing yield of the central focusing peak, and a decrease in the spacing between the peaks near the center.

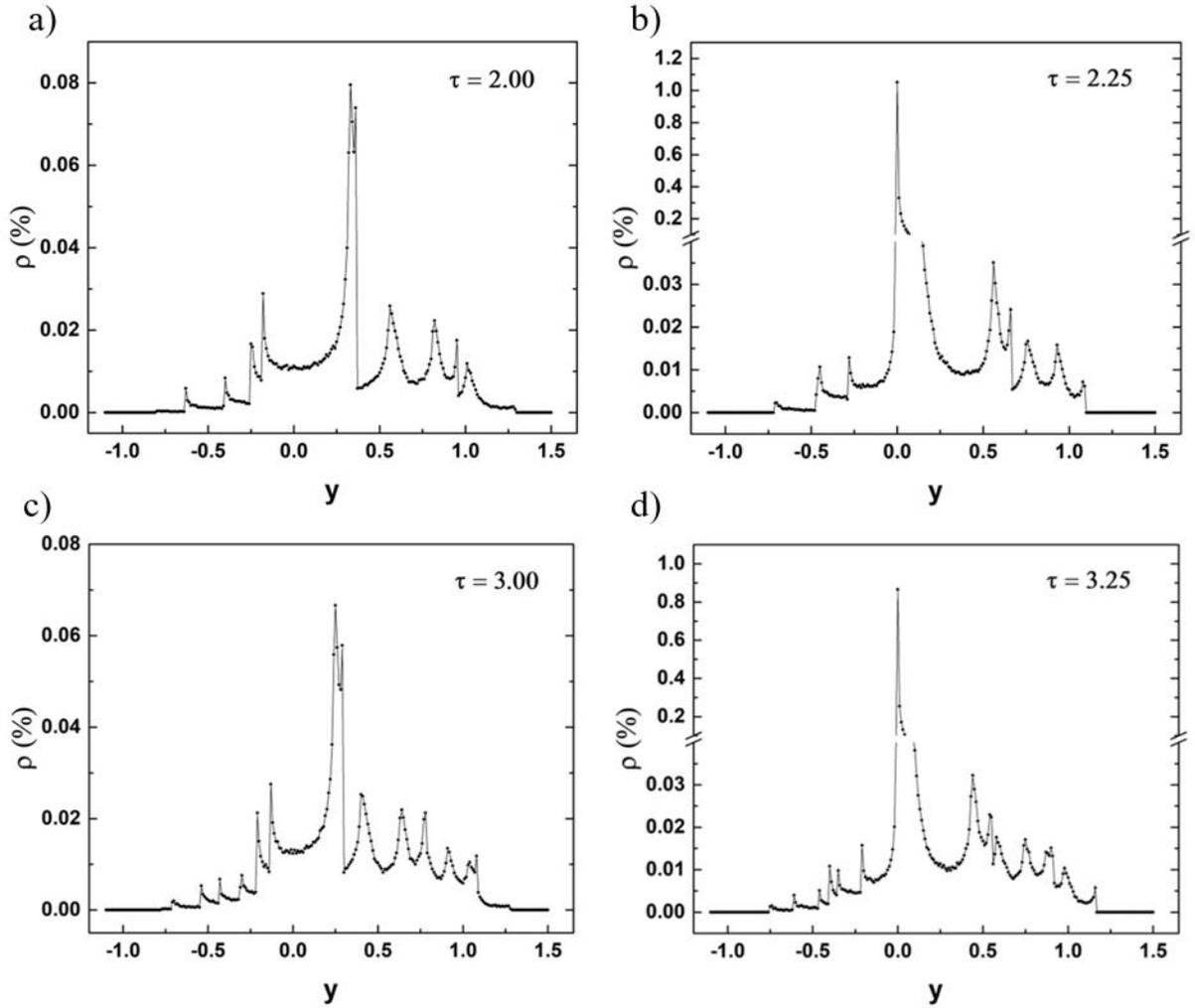

**Figure 3**. Cuts through the probability distributions for $x = 0$ at times equal to (a) $\tau = 2$, (b) $\tau = 2.25$, (c) $\tau = 3$, and (d) $\tau = 3.25$.

The Jacobian (5) for the Toda potential (3) can be determined numerically in the following way. First, the mapping (4) is obtained from the calculated particle trajectories. Second, the Jacobian is determined by using the high-precision numerical differentiation applying the Richardson extrapolation method of the six orders [29]. Solutions of the equation $J = 0$, corresponding to singularities, are obtained using the double quadratic spline interpolation [28]. It is important to note that calculating Jacobian requires tracking trajectories around an initial $(x_0, y_0)$ point. Consequently, the numerical procedure is dependable on time and accurate for small times only. For large times, the trajectories originated from the same initial point tend to drift apart, reducing the accuracy of numerical differentiation.

Figs. 4(a) and 4(b) depict the calculated singular patterns for $\tau = 2$ and 2.25, while Figs. 4(c) and 4(d) show the patterns for $\tau = 3$ and 3.25, respectively. For $\tau = 2$, the singularities consist of two double-cusp triangles and two lines parallel to the equipotential line $V = 1$. For $\tau = 3$, there are three double-cusp triangles and three lines parallel to the equipotential line $V = 1$. These singularities are in full accordance with the corresponding line-like structures presented in Figs. 2(a) and 2(c), respectively. Moreover, it is clear that they represent a "skeleton" of probability distribution. These singular patterns exhibit global structural stability, as they are of the fold and cusp type around all singular points.

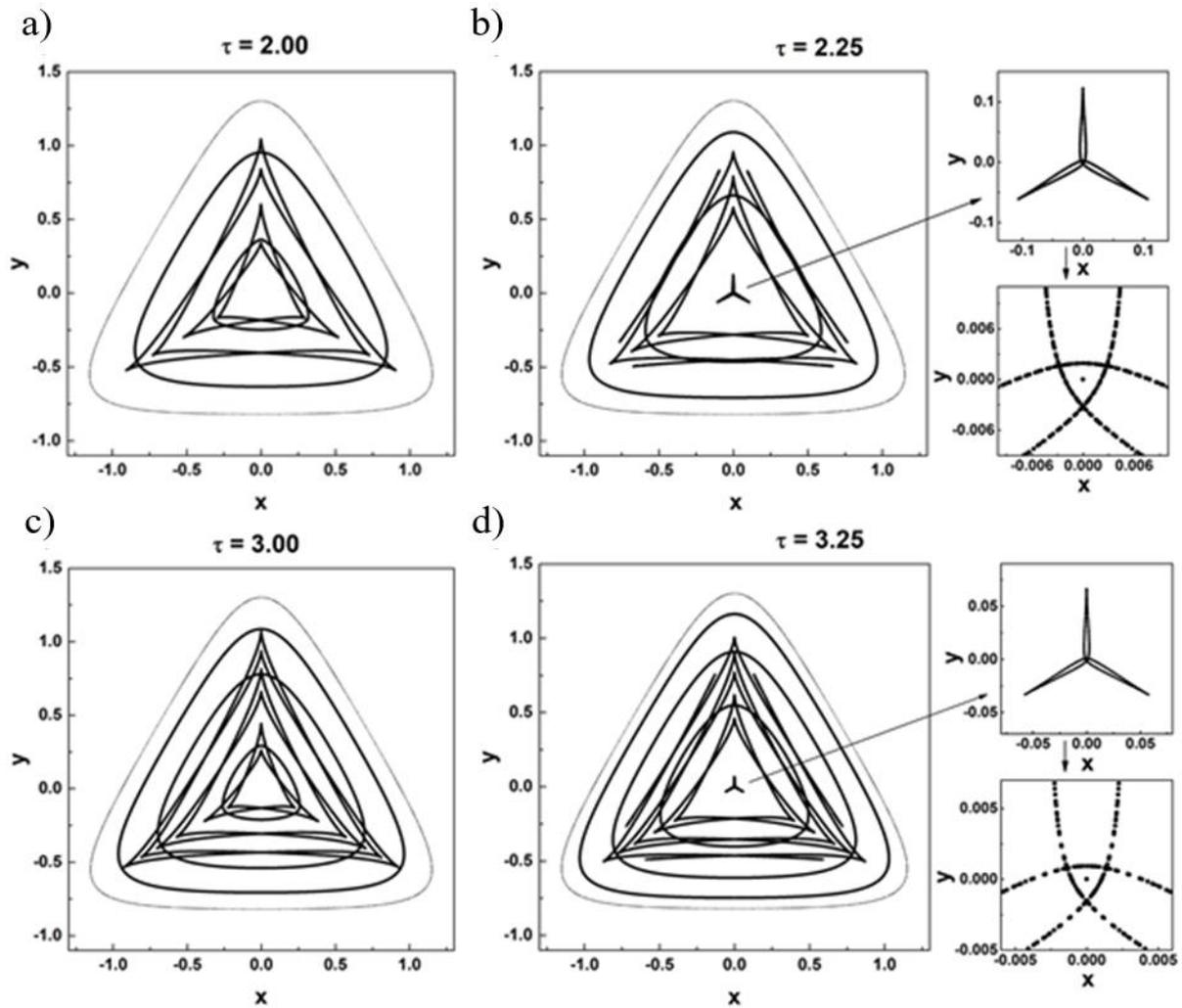

**Figure 4.** Singular patterns for $V < 1$ at times equal to (a) $\tau = 2$, (b) $\tau = 2.25$, (c) $\tau = 3$, and (d) $\tau = 3.25$. The line corresponding to $V = 1$ is designated by a thin solid line.

For $\tau = 2.25$, there are two singular lines parallel to the equipotential line $V = 1$, one double-cusped triangle, one single-cusped triangle and three lines attached to its sides. The analysis shows that they are, in fact, parts of a double-cusped triangle, whose cusps are not visible due to the fact that domain of the mapping (4) is restricted to $V(x, y) < 1$. Fig. 4(b) also shows zooms of the central singularities, revealing a propeller-like line in the upper right and a singular point in the lower right. Investigating the singularities for $V(x, y) \geq 1$ is beyond the scope of this work.

Fig. 4(d) shows singularities for $\tau = 3.25$ presented in the same way as for $\tau = 2.25$. Comparing the singularities for $\tau = 3.25$ with those for $\tau = 2.25$, one can observe an additional double-cusped triangle and a line parallel to the equipotential line $V = 1$, along with the same type of singularities around the center. Three lines attached to sides of the one-cusped triangle can be also observed. The analysis shows that they have the same origin as for $\tau = 2.25$. Therefore, for $\tau = 2.25$ and $3.25$, the pronounced peaks at the center are the singular peaks corresponding to singular points. As mentioned earlier, the numerical uncertainty in calculating the singular patterns increases with time; therefore, the singular patterns for larger times are not numerically accessible in this work.

To examine the structural stability of the central singular points at $\tau = 2.25$ and $3.25$, we analyzed the time evolution of the surrounding patterns. Figs. 5(a) and (b) show these results for times near $\tau = 2.25 - \tau = 2.247$ and $2.253$, and near $\tau = 3.25 - \tau = 3.247$ and $3.253$, respectively. In both cases, cusped triangles are formed both before and after the singular points. This demonstrates that, unlike the isolated singular points in the harmonic case, these points are part of a continuous time evolution of structurally stable cusped triangles. We observe a slight increase in the size of these cusped triangles over time. This sequential formation can be compared to an elliptic umbilic unfolding catastrophe, considering that the unfolding parameter changes slightly and continuously over time [19]. Time dependence of the unfolding parameter, which is not predicted by standard catastrophe theory, aligns with our finding that the singular pattern morphology changes dynamically with time as the natural parameter.

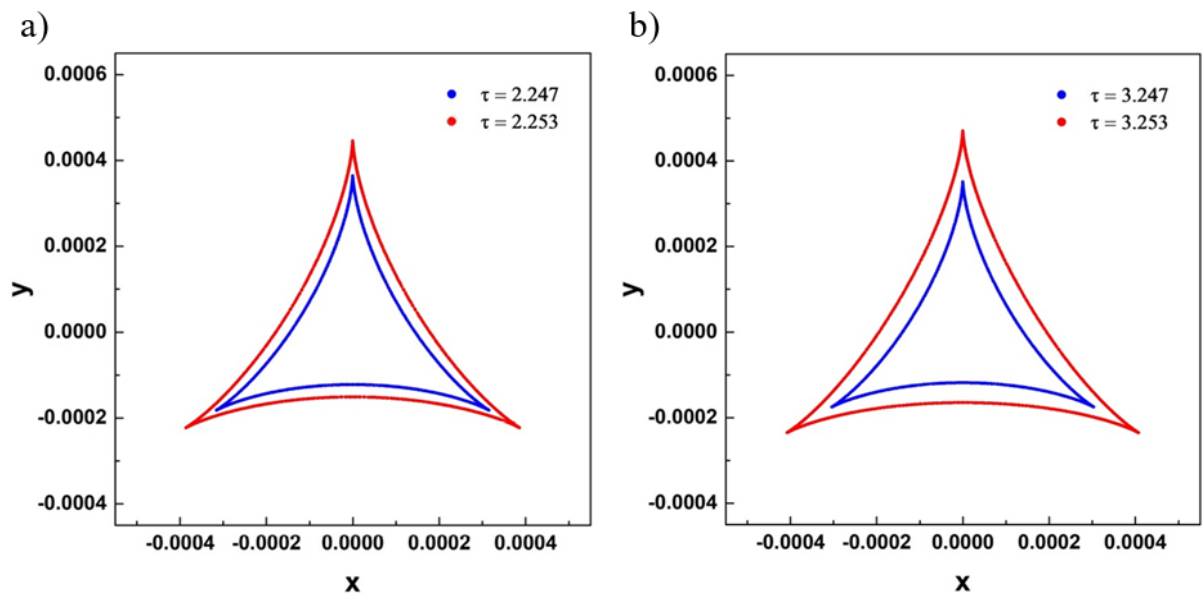

**Figure 5**. Singular patterns for times (a) $\tau = 2.247$ – blue color and $\tau = 2.253$ – red color and (b) $\tau = 3.247$ – blue color and $\tau = 3.253$ – red color.

From the previously presented results, it can be anticipated that the evolution of the probability distribution for $\tau = 2, 3, 4, ...$ is characterized by line-like structures corresponding to singular patterns that increase linearly, while their mutual distance decreases over time. Additionally, for $\tau = 2.25, 3.25, 4.25, ...$ the singular peaks occur at the center, corresponding to singular points. Their intensities decrease over time. This is demonstrated in Figs. 6(a) and 6(b) for $\tau = 10$ and $10.25$, respectively. The cuts of these distributions, shown in Figs. 6(c) and 6(d), confirm the trend of an increasing number of peaks, decreasing distances between them, and decreasing yield of the central peak. Additionally, the peaks surrounding the center also move closer together. This trend suggests that at a certain time, all singularities near the center will merge into a single central singular peak. The analysis shows that this event occurs at approximately $\tau = 100$.

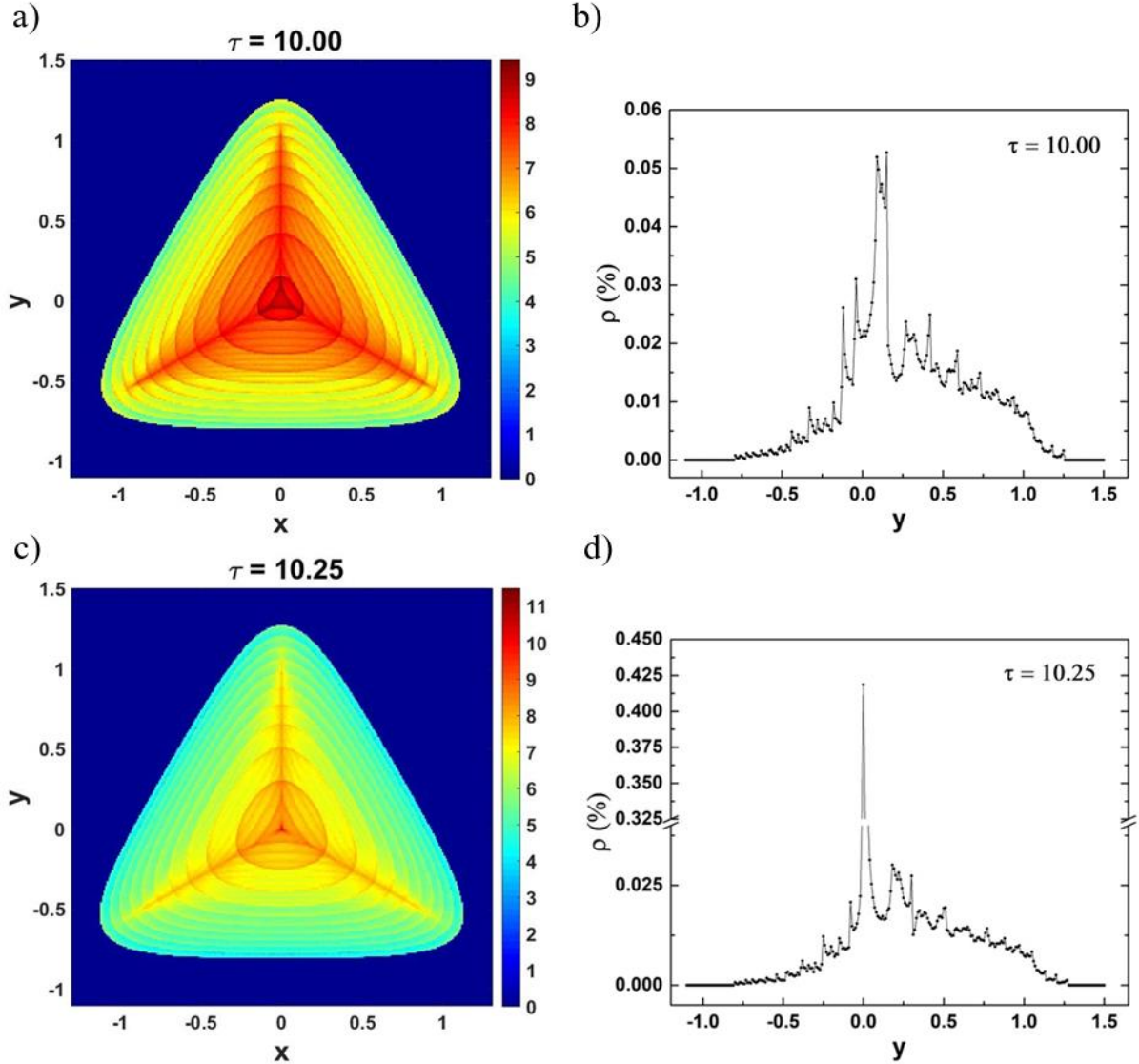

**Figure 6.** The 2D distribution at times equal to (a) $\tau = 10$ and (b) $\tau = 10.25$. The jet color map in a logarithmic scale is used. Cuts through the probability distributions for $x = 0$ at times equal to (c) $\tau = 10$, and (d) $\tau = 10.25$.

From the numerical point of view, a steady state probability distribution can be achieved when the yield in a bin remains effectively constant over time. Given that the number of double cusped-like triangles and singular lines parallel to the equipotential line $V = 1$ increase linearly, a total of approximately 1000 characteristic singular lines are required to reach a steady state. Namely, since the intersection of the line $x = 0$ and the equipotential line $V = 1$ has a length of 2.13, the average mutual distances between the singularities can be estimated to be $2.13/1000 = 0.0021 \ll 0.01$. Therefore, each bin would contain an average of five singular peaks, resulting in practically no observable structure within the bin.

Fig. 7(a) shows the probability distribution for a very large time, $\tau = 500$, corresponding to a total of 1000 characteristic singular lines. As predicted, the distribution is numerically smooth, with a pronounced peak at the center. Fig. 7(b) shows cuts through the probability distributions for $x = 0$ and times $\tau = 500$ and $\tau = 500.25$. Both curves presented in this figure are smooth and practically the same, except the yields in a very small area around the center with a diameter of 0.02. For $\tau = 500$, the central yield is equal to 0.088%, while for $\tau = 500.25$, the yield is equal to 0.096%. A small difference in yield near the center can be attributed to the dominance of harmonic oscillatory trajectories in this region, as the anharmonic component becomes negligible.

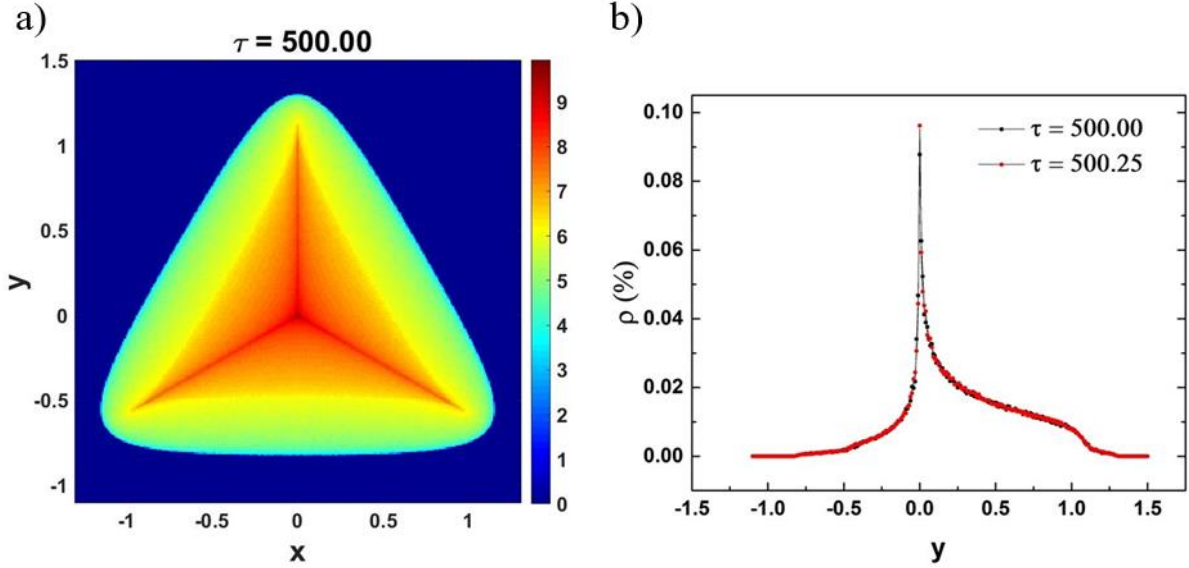

**Figure 7.** (a) The 2D distribution at time equal to $\tau = 500$. The jet color map in a logarithmic scale is used. (b) Cuts through the probability distributions for $x = 0$ at times equal to $\tau = 500$, and (d) $\tau = 500.25$.

Based on the observed trend that the central yield difference between $\tau = n\frac{1}{2}$ and $\tau = \frac{1}{4} + n\frac{1}{2}$ decreases with increasing $n$ it is anticipated that this difference approaches zero as $\tau \to \infty$. However, this assumption cannot be verified numerically since times, $\tau \to \infty$, are inaccessible. Moreover, increasing the initial number of particles and reducing the bin size would extend the time required to reach a steady state beyond $\tau = 500$. Nonetheless, *the singular patterns* will *ultimately drive the system toward a steady state as $\tau \to \infty$*.

**Conclusions**

The evolution of the probability distribution in the configuration space for the Toda potential exhibits regular yet complex behavior. The probability distribution is characterized by singular patterns that represent the "skeleton" of the distribution. The singular pattern mainly consists of double-cusped triangular lines and lines parallel to the equipotential line, $V = 1$. Their number increases linearly, and the average distance between them decreases with time. The singular web that fills the accessible area $V(x, y) < 1$ over time drives the probability distribution toward a steady state. Considering the applied statistics and the resolution of the probability distribution, the steady state is reached for $\tau \geq 500$. At this point, the distribution is numerically smooth, with a pronounced singular peak at the center. One can anticipate that similar behavior occurs in regions where $V(x, y) \geq 1$.

Further investigations are needed into other symmetric, bounded, two-dimensional conservative systems, with a focus on comparing integrable and non-integrable systems and examining the connection between singularities in both cases. It can be hypothesized that the singularities of the associated Hamiltonian flow play a critical role in shaping the probability distribution in such systems.

Another important area of exploration is the classical-quantum correspondence. The well-known fact that the square of the wave function's modulus defines the probability distribution of a quantum particle in the configuration space and that the continuity equation for the quantum probability distribution holds exactly raises an interesting question: What role do classical singularities play in the corresponding quantum case?

**Acknowledgments**

S.P. and N.S. acknowledge support by the Ministry of Science, Technological Development and Innovation of the Republic of Serbia (No. 451-03-66/2024-03/200017). L.H. acknowledges support by National Key R\&D Program of China under Grant No. 2023YFA1407100, by NSFC under Grants No. 12175090, No. 12247101 and by the 111 Project under Grant No. B20063.